\documentclass[
  nojss]{jss}

\usepackage[utf8]{inputenc}

\providecommand{\tightlist}{%
  \setlength{\itemsep}{0pt}\setlength{\parskip}{0pt}}

\author{
Emanuele Guidotti\\University of Neuchâtel
}
\title{\pkg{calculus}: High Dimensional Numerical and Symbolic Calculus
in \proglang{R}}

\Plainauthor{Emanuele Guidotti}
\Plaintitle{calculus: High Dimensional Numerical and Symbolic Calculus
in R}
\Shorttitle{\pkg{calculus}: High Dimensional Numerical and Symbolic
Calculus in \proglang{R}}

\Abstract{
The \proglang{R} package \pkg{calculus} implements \proglang{C++}
optimized functions for numerical and symbolic calculus, such as the
Einstein summing convention, fast computation of the Levi-Civita symbol
and generalized Kronecker delta, Taylor series expansion, multivariate
Hermite polynomials, high-order derivatives, ordinary differential
equations, differential operators and numerical integration in arbitrary
orthogonal coordinate systems. The library applies numerical methods
when working with \code{functions} or symbolic programming when working
with \code{characters} or \code{expressions}. The package handles
multivariate numerical calculus in arbitrary dimensions and coordinates
and implements the symbolic counterpart of the numerical methods
whenever possible, without depending on external computer algebra
systems. Except for \pkg{Rcpp}, the package has no strict dependencies
in order to provide a stable self-contained toolbox that invites re-use.
}

\Plainkeywords{symbolic programming, finite difference, differential
operators, numerical integration, coordinate systems, Einstein
summation, Taylor series, Hermite polynomials, R}

\Address{
    Emanuele Guidotti\\
  University of Neuchâtel\\
  Institute of Financial Analysis University of Neuchâtel Rue
Abram-Louis-Breguet 2, 2000 Neuchâtel, Switzerland.\\
  E-mail: \email{emanuele.guidotti@unine.ch}\\
  
  }

\usepackage{amsmath} \usepackage{amsfonts} \usepackage{booktabs} \usepackage{longtable}

\begin{document}

\hypertarget{introduction}{%
\section{Introduction}\label{introduction}}

Multivariate calculus underlies a wide range of applications in the
natural and social sciences. In statistics, asymptotic expansion
formulas for stochastic processes \citep{yoshida1992asymptotic} can be
obtained by solving high dimensional systems of ordinary differential
equations. The transition density of multivariate diffusions can be
approximated using Hermite polynomials \citep{ait2002maximum} or
Taylor-like expansions \citep{li2013maximum}. Advances in medical
imaging technology as well as telecommunication data-collection have
ushered in massive datasets that make multidimensional data more
commonplace \citep{li2018rtensor} and tensors - multidimensional arrays
- have recently become ubiquitous in signal and data analytics at the
confluence of signal processing, statistics, data mining, and machine
learning \citep{sidiropoulos2017tensor}. In Earth sciences, cartography,
quantum mechanics, relativity, and engineering, non-Cartesian
coordinates are often chosen to match the symmetry of the problem in
two, three and higher dimensions.

\proglang{R} \citep{r} has shown to be a viable computing environment
for implementing and applying numerical methods\footnote{\url{https://cloud.r-project.org/web/views/NumericalMathematics.html}}
as a practical tool for applied statistics. However, such methods are
seldom flexible enough to handle multivariate calculus in arbitrary
dimensions and coordinates. The package \pkg{numDeriv} \citep{numDeriv}
sets the standard for numerical differentiation in \proglang{R},
providing numerical gradients, Jacobians, and Hessians, but does not
support higher order derivatives or differentiation of tensor-valued
functions. \pkg{tensorA} \citep{tensorA} implements the Einstein summing
convention but does not support arbitrary expressions involving more
than two tensors or tensors with repeated indices. \pkg{mpoly}
\citep{kahle2013mpoly} implements univariate but not multivariate
Hermite polynomials. In a similar way, \pkg{pracma} \citep{pracma}
supports the computation of Taylor series for univariate but not
multivariate functions. \pkg{cubature} \citep{cubature} provides an
efficient interface for multivariate integration but limited to
Cartesian coordinates.

On the other hand, \proglang{R} is not designed for symbolic computing.
Nevertheless, the advent of algebraic statistics and its contributions
to asymptotic theory in statistical models, experimental design,
multiway contingency tables, and disclosure limitation has increased the
need for \proglang{R} to be able to do some relatively basic operations
and routines with multivariate symbolic calculus \citep{kahle2013mpoly}.
Although there exist packages to interface external computer algebra
systems, \proglang{R} still lacks a native support that invites re-use.
The package \pkg{Ryacas} \citep{Ryacas} interfaces the computer algebra
system \emph{Yacas}\footnote{\url{http://www.yacas.org}}, while
\pkg{caracas} \citep{caracas} - based on \pkg{reticulate}
\citep{reticulate} - and \pkg{rSymPy} \citep{rSymPy} - based on
\pkg{rJython} \citep{rJython} - both access the symbolic algebra system
\emph{SymPy}\footnote{\url{http://www.sympy.org}}.

This work presents the \proglang{R} package \pkg{calculus} for high
dimensional numerical and symbolic calculus in \proglang{R}. The
contribution is twofold. First, the package handles multivariate
numerical calculus in arbitrary dimensions and coordinates via
\proglang{C++} optimized functions, improving the state-of-the-art both
in terms of flexibility and efficiency. It achieves approximately the
same accuracy for numerical differentiation as the \pkg{numDeriv}
\citep{numDeriv} package but significantly reduces the computational
time. It supports higher order derivatives and the differentiation of
possibly tensor-valued functions. Differential operators such as the
gradient, divergence, curl, and Laplacian are made available in
arbitrary orthogonal coordinate systems. The Einstein summing convention
supports expressions involving more than two tensors and tensors with
repeated indices. Besides being more flexible, the summation proves to
be faster than the alternative implementation found in the \pkg{tensorA}
\citep{tensorA} package for advanced tensor arithmetic with named
indices. Unlike \pkg{mpoly} \citep{kahle2013mpoly} and \pkg{pracma}
\citep{pracma}, the package supports multidimensional Hermite
polynomials and Taylor series of multivariate functions. The package
integrates seamlessly with \pkg{cubature} \citep{cubature} for efficient
numerical integration in \proglang{C} and extends the numerical
integration to arbitrary orthogonal coordinate systems. Second, the
symbolic counterpart of the numerical methods are implemented whenever
possible to meet the growing needs for \proglang{R} to handle basic
symbolic operations. The package provides, among others, symbolic high
order derivatives of possibly tensor-valued functions, symbolic
differential operators in arbitrary orthogonal coordinate systems,
symbolic Einstein summing convention and Taylor series expansion of
multivariate functions. This is done entirely in \proglang{R}, without
depending on external computer algebra systems in order to provide a
self-contained toolbox that invites re-use.

The remainder of the paper is organized as follows: Section
\ref{sec:calculus} introduces the package and the underlying philosophy,
Section \ref{sec:vector} and \ref{sec:matrix} provide basic utilities
for vector and matrix algebra, Section \ref{sec:tensor} presents tensor
algebra with particular focus on the Einstein summation, Section
\ref{sec:derivatives} provides fast and accurate derivatives, Section
\ref{sec:taylor} presents the Taylor series of possibly multivariate
functions, Section \ref{sec:hermite} describes multidimensional Hermite
polynomials, Section \ref{sec:ode} solves ordinary differential
equations, Section \ref{sec:differentials} and \ref{sec:integrals}
introduce differential operators and integrals in arbitrary orthogonal
coordinate systems before Section \ref{sec:summary} concludes.

\hypertarget{the-r-package-calculus}{%
\section{The R package calculus}\label{the-r-package-calculus}}

\label{sec:calculus}

The \proglang{R} package \pkg{calculus} implements \proglang{C++}
optimized functions for numerical and symbolic calculus, such as the
Einstein summing convention, fast computation of the Levi-Civita symbol
and generalized Kronecker delta, Taylor series expansion, multivariate
Hermite polynomials, high-order derivatives, ordinary differential
equations, differential operators and numerical integration in arbitrary
orthogonal coordinate systems.

\hypertarget{testing}{%
\subsection{Testing}\label{testing}}

Several unit tests are implemented via the standard framework offered by
\pkg{testthat} \citep{testthat} and run via continuous integration on
Travis CI\footnote{\url{https://travis-ci.com}}.

\hypertarget{dependencies}{%
\subsection{Dependencies}\label{dependencies}}

The package integrates seamlessly with \pkg{cubature} \citep{cubature}
for efficient numerical integration in \proglang{C}. However, except for
\pkg{Rcpp} \citep{eddelbuettel2011rcpp}, the package has no strict
dependencies in order to provide a stable self-contained toolbox that
invites re-use.

\hypertarget{installation}{%
\subsection{Installation}\label{installation}}

The stable release version of \pkg{calculus} is hosted on the
Comprehensive R Archive Network (CRAN) at
\url{https://CRAN.R-project.org/package=calculus} and it can be
installed using:

\begin{CodeChunk}

\begin{CodeInput}
R> install.packages("calculus")
\end{CodeInput}
\end{CodeChunk}

\hypertarget{philosophy}{%
\subsection{Philosophy}\label{philosophy}}

The package provides a unified interface to work with mathematical
objects in \proglang{R}. The library applies numerical methods when
working with \code{functions} or symbolic programming when working with
\code{characters} or \code{expressions}. To describe multidimensional
objects such as vectors, matrices, and tensors, the package uses the
class \code{array} regardless of the dimension. This is done to prevent
unwanted results due to operations among different classes such as
\code{vector} for unidimensional objects or \code{matrix} for
bidimensional objects.

\hypertarget{basic-arithmetic}{%
\subsection{Basic arithmetic}\label{basic-arithmetic}}

Basic elementwise operations are supported for arrays of the same
dimensions with the following data types: \code{numeric},
\code{complex}, \code{character}, \code{expression}. Automatic type
conversion is supported and string manipulation is performed in
\proglang{C++} to improve performance. A minimal simplification
algorithm is included to simplify operations involving zeros. Below a
unidimensional example on the sum, difference, product, and division
among the different data types.

\begin{CodeChunk}

\begin{CodeInput}
R> ("a + b" %prod% 1i) %sum% (0 %prod% "c") %diff% (expression(d+e) %div% 3)
\end{CodeInput}

\begin{CodeOutput}
[1] "((a + b) * (0+1i)) - ((d + e) / 3)"
\end{CodeOutput}
\end{CodeChunk}

The characters are automatically wrapped in parentheses when performing
basic symbolic operations to prevent unwanted results, e.g.:

\[a+b \cdot c+d\]

instead of

\[(a+b) \cdot (c+d)\]

To disable this behaviour the user can set
\code{options(calculus.auto.wrap = FALSE)}.

\hypertarget{vector-algebra}{%
\section{Vector algebra}\label{vector-algebra}}

\label{sec:vector}

A vector can be regarded as a 1-dimensional tensor. In \proglang{R}, it
can be regarded as a 1-dimensional \code{array} so that the methods
presented for multidimensional tensors in Section \ref{sec:tensor} are
available for vectors. The package also implements a few vector-specific
utilities, such as the cross product.

\hypertarget{cross-product}{%
\subsection{Cross product}\label{cross-product}}

The cross product or vector product is an operation on \(n-1\) vectors
in \(n\)-dimensional space. The results is a \(n\)-dimensional vector
that is perpendicular to the \(n-1\) vectors. For example in
\(\mathbb{R}^3\):

\begin{CodeChunk}

\begin{CodeInput}
R> cross(c(1,0,0), c(0,1,0))
\end{CodeInput}

\begin{CodeOutput}
[1] 0 0 1
\end{CodeOutput}
\end{CodeChunk}

And in \(\mathbb{R}^4\):

\begin{CodeChunk}

\begin{CodeInput}
R> cross(c(1,0,0,0), c(0,1,0,0), c(0,0,0,1))
\end{CodeInput}

\begin{CodeOutput}
[1] 0 0 1 0
\end{CodeOutput}
\end{CodeChunk}

Consistently with the philosophy of the package, the same interface is
provided for \code{character} vectors.

\begin{CodeChunk}

\begin{CodeInput}
R> cross(c("a","b","c"), c("d","e","f"))
\end{CodeInput}

\begin{CodeOutput}
[1] "((b)*((f)) + -(e)*((c))) * 1"  "((a)*((f)) + -(d)*((c))) * -1"
[3] "((a)*((e)) + -(d)*((b))) * 1" 
\end{CodeOutput}
\end{CodeChunk}

\hypertarget{matrix-algebra}{%
\section{Matrix algebra}\label{matrix-algebra}}

\label{sec:matrix}

A matrix can be regarded as a 2-dimensional tensor. In \proglang{R}, it
can be regarded as a 2-dimensional \code{array} so that the methods
presented for multidimensional tensors in Section \ref{sec:tensor} are
available for matrices. The package also implements a few
matrix-specific utilities, such as the symbolic determinant, inverse,
and matrix product.

\hypertarget{determinant}{%
\subsection{Determinant}\label{determinant}}

\label{sec:determinant}

The function \code{mxdet} computes the numerical or symbolic determinant
of matrices depending on the data type. If the elements of the
\code{matrix} are of type \code{numeric}, then the determinant is
computed via the function \code{det} available in base \proglang{R}.

\begin{CodeChunk}

\begin{CodeInput}
R> mxdet(matrix(1:4, nrow = 2))
\end{CodeInput}

\begin{CodeOutput}
[1] -2
\end{CodeOutput}
\end{CodeChunk}

If the elements are of type \code{character}, then the symbolic
determinant is computed recursively in \proglang{C++}.

\begin{CodeChunk}

\begin{CodeInput}
R> mxdet(matrix(letters[1:4], nrow = 2))
\end{CodeInput}

\begin{CodeOutput}
[1] "a*(d) + -b*(c)"
\end{CodeOutput}
\end{CodeChunk}

The symbolic determinant offers a significant gain in performance when
computing determinants for a large number of matrices. The following
test compares the performance of two different approaches to compute the
determinant of \(2^{16}\) 4x4-matrices. Method \emph{numeric}: compute
the numeric determinant for each matrix. Method \emph{symbolic}: compute
the symbolic determinant of a 4x4-matrix and evaluate it for each
matrix.

\begin{CodeChunk}

\begin{CodeInput}
R> n <- 4
R> e <- letters[1:n^2]
R> grid <- expand.grid(lapply(1:n^2, function(e) runif(2)))
R> colnames(grid) <- e
R> microbenchmark(
R+   "numeric" = {
R+     x <- apply(grid, 1, function(e) det(matrix(e, nrow = n)))
R+   },
R+   "symbolic" = {
R+     x <- evaluate(c2e(mxdet(matrix(e, nrow = n))), grid)   
R+   }
R+ )
\end{CodeInput}

\begin{CodeOutput}
Unit: milliseconds
     expr      min       lq   mean   median      uq      max neval cld
  numeric 911.6423 935.7297 967.09 964.9941 993.513 1086.197   100   b
 symbolic   7.4456   8.1825  10.44   9.2427  13.423   16.848   100  a 
\end{CodeOutput}
\end{CodeChunk}

\hypertarget{matrix-inverse}{%
\subsection{Matrix Inverse}\label{matrix-inverse}}

The function \code{mxinv} computes the numerical or symbolic inverse of
matrices depending on the data type. If the elements of the
\code{matrix} are of type \code{numeric}, then the inverse is computed
via the function \code{solve} available in base \proglang{R}.

\begin{CodeChunk}

\begin{CodeInput}
R> mxinv(matrix(1:4, byrow = TRUE, nrow = 2))
\end{CodeInput}

\begin{CodeOutput}
     [,1] [,2]
[1,] -2.0  1.0
[2,]  1.5 -0.5
\end{CodeOutput}
\end{CodeChunk}

If the elements are of type \code{character}, then the symbolic inverse
is computed based on the determinants in Section \ref{sec:determinant}.

\begin{CodeChunk}

\begin{CodeInput}
R> mxinv(matrix(letters[1:4], byrow = TRUE, nrow = 2))
\end{CodeInput}

\begin{CodeOutput}
     [,1]                      [,2]                     
[1,] "(d) / (a*(d) + -c*(b))"  "-(b) / (a*(d) + -c*(b))"
[2,] "-(c) / (a*(d) + -c*(b))" "(a) / (a*(d) + -c*(b))" 
\end{CodeOutput}
\end{CodeChunk}

The symbolic inverse offers a gain in performance when inverting a large
number of matrices, as shown by replicating the test in section
\ref{sec:determinant} and replacing the determinant with the inverse.

\begin{CodeChunk}

\begin{CodeInput}
R> n <- 4
R> e <- letters[1:n^2]
R> grid <- expand.grid(lapply(1:n^2, function(e) runif(2)))
R> colnames(grid) <- e
R> microbenchmark(
R+   "numeric" = {
R+     x <- apply(grid, 1, function(e) solve(matrix(e, nrow = n)))
R+   },
R+   "symbolic" = {
R+     x <- evaluate(c2e(mxinv(matrix(e, nrow = n))), grid)   
R+   }
R+ )
\end{CodeInput}

\begin{CodeOutput}
Unit: milliseconds
     expr    min      lq    mean  median      uq     max neval cld
  numeric 1507.0 1625.25 1693.77 1702.14 1732.92 2066.72   100   b
 symbolic  192.5  224.77  279.43  249.18  342.02  469.23   100  a 
\end{CodeOutput}
\end{CodeChunk}

\hypertarget{matrix-product}{%
\subsection{Matrix Product}\label{matrix-product}}

The matrix product can be expressed in Einstein notation as shown in
Section \ref{sec:tensor}, thus inheriting the support for symbolic
calculations.

\begin{CodeChunk}

\begin{CodeInput}
R> a <- matrix(1:4, nrow = 2, byrow = TRUE)
R> b <- matrix(letters[1:4], nrow = 2, byrow = TRUE)
R> a %mx% b
\end{CodeInput}

\begin{CodeOutput}
     [,1]                [,2]               
[1,] "1 * (a) + 2 * (c)" "1 * (b) + 2 * (d)"
[2,] "3 * (a) + 4 * (c)" "3 * (b) + 4 * (d)"
\end{CodeOutput}
\end{CodeChunk}

\hypertarget{tensor-algebra}{%
\section{Tensor algebra}\label{tensor-algebra}}

\label{sec:tensor}

A tensor may be represented as a multidimensional array. Just as a
vector in an \(n\)-dimensional space is represented by a
\(1\)-dimensional array with \(n\) components, a matrix is represented
by a \(2\)-dimensional array with \(n_1\;\text{x}\;n_2\) components, and
any tensor can be represented by a \(d\)-dimensional array with
\({n_1\;\text{x}\;\dots\; \text{x}\; n_d}\) components. This makes the
class \code{array} available in base \proglang{R} an ideal candidate to
represent mathematical tensors. In particular, the class stores its
dimensions in the attribute \code{dim} that contains a vector giving the
length for each dimension.

\begin{CodeChunk}

\begin{CodeInput}
R> A <- array(1:24, dim = c(2,3,4))
R> attributes(A)
\end{CodeInput}

\begin{CodeOutput}
$dim
[1] 2 3 4
\end{CodeOutput}
\end{CodeChunk}

The package \pkg{calculus} reads this attribute to represent tensors in
index notation, such as \(A_{ijk}\). In particular, the function
\code{index} is used to assign indices to the dimensions of the tensor
by setting names to the attribute \code{dim}.

\begin{CodeChunk}

\begin{CodeInput}
R> index(A) <- c("i","j","k")
R> attributes(A)
\end{CodeInput}

\begin{CodeOutput}
$dim
i j k 
2 3 4 
\end{CodeOutput}
\end{CodeChunk}

In this way, a tensor with named indices is represented by an
\code{array} with named \code{dim}. The package \pkg{calculus} builds a
set of tools to work with tensors on top of this class and provides the
implementation of the Levi-Civita symbol and Generalized Kronecker delta
that often appears in tensor algebra. At the time of writing, the
package makes no distinction between upper and lower indices,
i.e.~vectors and covectors\footnote{\url{https://en.wikipedia.org/wiki/Einstein_notation}}.

\hypertarget{levi-civita-symbol}{%
\subsubsection{Levi-Civita symbol}\label{levi-civita-symbol}}

In mathematics, particularly in linear algebra, tensor analysis, and
differential geometry, the Levi-Civita symbol represents a collection of
numbers; defined from the sign of a permutation of the natural numbers
\(1, 2, …, n\), for some positive integer \(n\). It is named after the
Italian mathematician and physicist Tullio Levi-Civita. Other names
include the permutation symbol, antisymmetric symbol, or alternating
symbol, which refer to its antisymmetric property and definition in
terms of permutations.\footnote{\url{https://en.wikipedia.org/wiki/Levi-Civita_symbol}}
In the general \(n\)-dimensional case, the Levi-Civita symbol is defined
by:

\[
\varepsilon _{i_{1}i_{2}\dots i_{n}}={\begin{cases}+1&{\text{if }}(i_{1},i_{2},\ldots ,i_{n}){\text{ is an even permutation of }}(1,2,\dots ,n)\\-1&{\text{if }}(i_{1},i_{2},\dots ,i_{n}){\text{ is an odd permutation of }}(1,2,\dots ,n)\\\;\;\,0&{\text{otherwise}}\end{cases}}
\]

The function \code{epsilon} determines the parity of the permutation in
\proglang{C++} via efficient cycle decomposition\footnote{\url{https://www.geeksforgeeks.org/number-of-transpositions-in-a-permutation/}}
and constructs the Levi-Civita symbol in arbitrary dimension. For
example the 2-dimensional Levi-Civita symbol is given by:

\[
\varepsilon _{ij}={\begin{cases}+1&{\text{if }}(i,j)=(1,2)\\-1&{\text{if }}(i,j)=(2,1)\\\;\;\,0&{\text{if }}i=j\end{cases}}
\]

\begin{CodeChunk}

\begin{CodeInput}
R> epsilon(2)
\end{CodeInput}

\begin{CodeOutput}
     [,1] [,2]
[1,]    0    1
[2,]   -1    0
\end{CodeOutput}
\end{CodeChunk}

And in 3 dimensions:

\[
\varepsilon _{ijk}={\begin{cases}+1&{\text{if }}(i,j,k){\text{ is }}(1,2,3),(2,3,1),{\text{ or }}(3,1,2),\\-1&{\text{if }}(i,j,k){\text{ is }}(3,2,1),(1,3,2),{\text{ or }}(2,1,3),\\\;\;\,0&{\text{if }}i=j,{\text{ or }}j=k,{\text{ or }}k=i\end{cases}}
\]

\begin{CodeChunk}

\begin{CodeInput}
R> epsilon(3)
\end{CodeInput}

\begin{CodeOutput}
, , 1

     [,1] [,2] [,3]
[1,]    0    0    0
[2,]    0    0    1
[3,]    0   -1    0

, , 2

     [,1] [,2] [,3]
[1,]    0    0   -1
[2,]    0    0    0
[3,]    1    0    0

, , 3

     [,1] [,2] [,3]
[1,]    0    1    0
[2,]   -1    0    0
[3,]    0    0    0
\end{CodeOutput}
\end{CodeChunk}

\hypertarget{generalized-kronecker-delta}{%
\subsubsection{Generalized Kronecker
delta}\label{generalized-kronecker-delta}}

The generalized Kronecker delta or multi-index Kronecker delta of order
\(2p\) is a type \((p,p)\) tensor that is a completely antisymmetric in
its \(p\) upper indices, and also in its \(p\) lower indices.\footnote{\url{https://en.wikipedia.org/wiki/Kronecker_delta}}
In terms of the indices, the generalized Kronecker delta is defined as
\citep{frankel2011geometry}:

\[
\delta _{\nu _{1}\dots \nu _{p}}^{\mu _{1}\dots \mu _{p}}={\begin{cases}+1&\quad {\text{if }}(\nu _{1}\dots \nu _{p}){\text{ is an even permutation of }}(\mu _{1}\dots \mu _{p})\\-1&\quad {\text{if }}(\nu _{1}\dots \nu _{p}){\text{ is an odd permutation of }}(\mu _{1}\dots \mu _{p})\\\;\;0&\quad {\text{otherwise}}\end{cases}}
\]

When \(p=1\), the definition reduces to the standard Kronecker delta
that corresponds to the \(n\) x \(n\) identity matrix
\(I_{ij}=\delta^i_j\) where \(i\) and \(j\) take the values
\(1, 2, \dots, n\).

\begin{CodeChunk}

\begin{CodeInput}
R> delta(n = 3, p = 1)
\end{CodeInput}

\begin{CodeOutput}
     [,1] [,2] [,3]
[1,]    1    0    0
[2,]    0    1    0
[3,]    0    0    1
\end{CodeOutput}
\end{CodeChunk}

\hypertarget{tensor-contraction}{%
\subsection{Tensor contraction}\label{tensor-contraction}}

Tensor contraction can be seen as a generalization of the trace for a
square matrix. In the general case, a tensor can be contracted by
summing over pairs of repeated indices that share the same dimension.
This is achieved via the \proglang{C++} optimized function
\code{contraction}. Consider the following \(2\text{x}2\text{x}2\)
tensor:

\begin{CodeChunk}

\begin{CodeInput}
R> x <- array(1:8, dim = c(2,2,2))
R> print(x)
\end{CodeInput}

\begin{CodeOutput}
, , 1

     [,1] [,2]
[1,]    1    3
[2,]    2    4

, , 2

     [,1] [,2]
[1,]    5    7
[2,]    6    8
\end{CodeOutput}
\end{CodeChunk}

The trace of the tensor \(T=\sum_{i}T_{iii}\) is obtained with:

\begin{CodeChunk}

\begin{CodeInput}
R> contraction(x)
\end{CodeInput}

\begin{CodeOutput}
[1] 9
\end{CodeOutput}
\end{CodeChunk}

The contraction on the first and third dimension \(T_j=\sum_{i}T_{iji}\)
can be computed with:

\begin{CodeChunk}

\begin{CodeInput}
R> index(x) <- c("i","j","i")
R> contraction(x)
\end{CodeInput}

\begin{CodeOutput}
[1]  7 11
\end{CodeOutput}
\end{CodeChunk}

Finally, it is possible to preserve the dummy dimensions
\(T_{ij}=T_{iji}\) by setting the argument \code{drop = FALSE}:

\begin{CodeChunk}

\begin{CodeInput}
R> index(x) <- c("i","j","i")
R> contraction(x, drop = FALSE)
\end{CodeInput}

\begin{CodeOutput}
     [,1] [,2]
[1,]    1    6
[2,]    3    8
\end{CodeOutput}
\end{CodeChunk}

In this way, it is possible to compute arbitrary contraction of tensors
such as \(T_{klm}=\sum_{ij}T_{ikiiljjm}\) or \(T_{ijklm}=T_{ikiiljjm}\)
to preserve the dummy dimensions.

\hypertarget{einstein-summation}{%
\subsection{Einstein Summation}\label{einstein-summation}}

\label{sec:einstein}

In mathematics, the Einstein notation or Einstein summation convention
is a notational convention that implies summation over a set of repeated
indices. When an index variable appears twice, it implies summation over
all the values of the index\footnote{\url{https://en.wikipedia.org/wiki/Einstein_notation}}.
For instance the matrix product can be written in terms of Einstein
notation as:

\[C_{ij} = A_{ik}{B_{kj}}\equiv\sum_k A_{ik}{B_{kj}}\]

An arbitrary summation of the kind

\[D_{k}=A_{ijj}B_{iijk}C_{j}\equiv\sum_{ij}A_{ijj}B_{iijk}C_{j}=\sum_j\Bigl(\sum_{i}A_{ijj}B_{iijk}\Bigl)C_{j}\]

is implemented as follows:

\begin{enumerate}
\def\labelenumi{\arabic{enumi})}
\tightlist
\item
  Contract the first tensor and preserve the dummy dimensions:
  \(A_{ijj}\rightarrow A_{ij}\).
\item
  Contract the second tensor and preserve the dummy dimensions:
  \(B_{iijk}\rightarrow B_{ijk}\).
\item
  Permute and move the summation indices to the end:
  \(A_{ij}\rightarrow A_{ij}\), \(B_{ijk}\rightarrow B_{kij}\).
\item
  Compute the elementwise product on the repeated indices:
  \((AB)_{kij}=A_{ij}B_{kij}\).
\item
  Sum over the summation indices that do now appear in the other
  tensors: \[(AB)_{kj}=\sum_i(AB)_{kij}\]
\item
  Contract the third tensor and preserve the dummy dimensions:
  \(C_j\rightarrow C_j\).
\item
  Permute and move the summation indices to the end:
  \((AB)_{kj}\rightarrow (AB)_{kj}\), \(C_j\rightarrow C_j\).
\item
  Compute the elementwise product on the repeated indices:
  \((ABC)_{kj}=(AB)_{kj}C_{j}\).
\item
  Sum over the summation indices that do now appear in the other
  tensors: \[D_k = (ABC)_{k}=\sum_j(ABC)_{kj}\]
\item
  Iterate until all the tensors in the summation are considered.
\end{enumerate}

The function \code{einstein} provides a convenient way to compute
general Einstein summations among two or more tensors, with or without
repeated indices appearing in the same tensor. The function supports
both numerical and symbolical calculations implemented via the usage of
\proglang{C++} templates that operate with generic types and allow the
function to work on the different data types without being rewritten for
each one. The following example illustrates a sample Einstein summation
with mixed data types:

\[D_{jk}=A_{ij}B_{ki}C_{ii}\]

\begin{CodeChunk}

\begin{CodeInput}
R> A <- array(1:6, dim = c(i = 2, j = 3))
R> print(A)
\end{CodeInput}

\begin{CodeOutput}
     [,1] [,2] [,3]
[1,]    1    3    5
[2,]    2    4    6
\end{CodeOutput}
\end{CodeChunk}

\begin{CodeChunk}

\begin{CodeInput}
R> B <- array(1:4, dim = c(k = 2, i = 2))
R> print(B)
\end{CodeInput}

\begin{CodeOutput}
     [,1] [,2]
[1,]    1    3
[2,]    2    4
\end{CodeOutput}
\end{CodeChunk}

\begin{CodeChunk}

\begin{CodeInput}
R> C <- array(letters[1:4], dim = c(i = 2, i = 2))
R> print(C)
\end{CodeInput}

\begin{CodeOutput}
     [,1] [,2]
[1,] "a"  "c" 
[2,] "b"  "d" 
\end{CodeOutput}
\end{CodeChunk}

\begin{CodeChunk}

\begin{CodeInput}
R> einstein(A, B, C)
\end{CodeInput}

\begin{CodeOutput}
     [,1]                 [,2]                 
[1,] "1 * (a) + 6 * (d)"  "2 * (a) + 8 * (d)"  
[2,] "3 * (a) + 12 * (d)" "6 * (a) + 16 * (d)" 
[3,] "5 * (a) + 18 * (d)" "10 * (a) + 24 * (d)"
\end{CodeOutput}
\end{CodeChunk}

In the particular case of Einstein summations between two numeric
tensors that, after proper contraction and permutation, can be rewritten
as

\[C_{i_1\dots i_a,j_1\dots j_b}=A_{i_1\dots i_a,k_1\dots k_n}B_{k_1\dots k_n,j_1\dots j_b}\]

the function implements the following scheme:

\begin{enumerate}
\def\labelenumi{\arabic{enumi})}
\tightlist
\item
  Reshape the tensor \(A_{i_1\dots i_a,k_1\dots k_n}\) in the matrix
  \(A_{I,K}\) where the dimension of \(I\) is the product of the
  dimensions of \(i_1\dots i_a\) and the dimensions of \(K\) is the
  product of the dimensions of \(k_1\dots k_n\).
\item
  Reshape the tensor \(B_{k_1\dots k_n,j_1\dots j_b}\) in the matrix
  \(B_{K,J}\) where the dimension of \(K\) is the product of the
  dimensions of \(k_1\dots k_n\) and the dimensions of \(J\) is the
  product of the dimensions of \(j_1\dots j_b\).
\item
  Compute the matrix product \(C_{IJ}=A_{IK}B_{KJ}\).
\item
  Reshape the matrix \(C_{IJ}\) in the tensor
  \(C_{i_1\dots i_a,j_1\dots j_b}\).
\end{enumerate}

In this way, it is sufficient to change the attribute \code{dim} of the
\code{arrays} and the Einstein summation is written in terms of a matrix
product that can be computed efficiently in base \proglang{R}. This
approach is almost twice as fast as the alternative implementation for
the Einstein summation in the \proglang{R} package \pkg{tensorA} for
advanced tensor arithmetic with named indices \citep{tensorA}.

\begin{CodeChunk}

\begin{CodeInput}
R> a <- array(1:1000000, dim = c(a=2, i=5, j=100, k=50, d=20))
R> b <- array(1:100000, dim = c(a=2, j=100, i=5, l=100))
R> 
R> Ta <- tensorA::to.tensor(a)
R> Tb <- tensorA::to.tensor(b)
R> 
R> microbenchmark(
R+   "calculus" = calculus::einstein(a, b),
R+   "tensorA" = tensorA::einstein.tensor(Ta, Tb)
R+ )
\end{CodeInput}

\begin{CodeOutput}
Unit: milliseconds
     expr     min     lq    mean  median      uq    max neval cld
 calculus  87.165  89.85  95.212  91.716  96.103 207.46   100  a 
  tensorA 146.204 152.58 170.172 157.642 169.796 274.36   100   b
\end{CodeOutput}
\end{CodeChunk}

\hypertarget{inner-product}{%
\subsection{Inner Product}\label{inner-product}}

The inner product is computed in base \proglang{R} for \code{numeric}
arrays or via Einstein summation for \code{character} arrays:

\[A_{i_1\dots i_n}B_{i_1\dots i_n}\]

\begin{CodeChunk}

\begin{CodeInput}
R> 1:3 %inner% letters[1:3]
\end{CodeInput}

\begin{CodeOutput}
[1] "1 * (a) + 2 * (b) + 3 * (c)"
\end{CodeOutput}
\end{CodeChunk}

\hypertarget{dot-product}{%
\subsubsection{Dot Product}\label{dot-product}}

The dot product between arrays with different dimensions is computed by
taking the inner product on the last dimensions of the two arrays. It is
written in Einstein notation as:

\[A_{i_1\dots i_aj_1\dots j_n}B_{j_1\dots j_n}\]

\begin{CodeChunk}

\begin{CodeInput}
R> matrix(1:6, byrow = TRUE, nrow = 2, ncol = 3) %dot% letters[1:3]
\end{CodeInput}

\begin{CodeOutput}
[1] "1 * (a) + 2 * (b) + 3 * (c)" "4 * (a) + 5 * (b) + 6 * (c)"
\end{CodeOutput}
\end{CodeChunk}

\hypertarget{outer-product}{%
\subsection{Outer Product}\label{outer-product}}

The outer product is computed in base \proglang{R} for \code{numeric}
arrays or via Einstein summation for \code{character} arrays:

\[A_{i_1\dots i_a}B_{j_1\dots j_b}\]

\begin{CodeChunk}

\begin{CodeInput}
R> 1:3 %outer% letters[1:3]
\end{CodeInput}

\begin{CodeOutput}
     [,1]      [,2]      [,3]     
[1,] "1 * (a)" "1 * (b)" "1 * (c)"
[2,] "2 * (a)" "2 * (b)" "2 * (c)"
[3,] "3 * (a)" "3 * (b)" "3 * (c)"
\end{CodeOutput}
\end{CodeChunk}

\hypertarget{kronecker-product}{%
\subsection{Kronecker Product}\label{kronecker-product}}

The package extends the generalized \code{kronecker} product available
in base \proglang{R} with support for arrays of type \code{character}.

\begin{CodeChunk}

\begin{CodeInput}
R> 1:3 %kronecker% letters[1:3]
\end{CodeInput}

\begin{CodeOutput}
[1] "1 * (a)" "1 * (b)" "1 * (c)" "2 * (a)" "2 * (b)" "2 * (c)" "3 * (a)"
[8] "3 * (b)" "3 * (c)"
\end{CodeOutput}
\end{CodeChunk}

\hypertarget{derivatives}{%
\section{Derivatives}\label{derivatives}}

\label{sec:derivatives}

The function \code{derivative} performs high-order symbolic and
numerical differentiation for generic tensors with respect to an
arbitrary number of variables. The function behaves differently
depending on the arguments \code{order}, the order of differentiation,
and \code{var}, the variable names with respect to which the derivatives
are computed.

When multiple variables are provided and \code{order} is a single
integer \(n\), then the \(n\)-th order derivative is computed for each
element of the tensor with respect to each variable:

\[D = \partial^{(n)} \otimes F\]

that is:

\[D_{i,\dots,j,k} = \partial^{(n)}_{k} F_{i,\dots,j}\]

where \(F\) is the tensor of functions and \(\partial_k^{(n)}\) denotes
the \(n\)-th order partial derivative with respect to the \(k\)-th
variable.

When \code{order} matches the length of \code{var}, it is assumed that
the differentiation order is provided for each variable. In this case,
each element is derived \(n_k\) times with respect to the \(k\)-th
variable, for each of the \(m\) variables.

\[D_{i,\dots,j} = \partial^{(n_1)}_1\cdots\partial^{(n_m)}_m F_{i,\dots,j}\]

The same applies when \code{order} is a named vector giving the
differentiation order for each variable. For example,
\code{order = c(x=1, y=2)} differentiates once with respect to \(x\) and
twice with respect to \(y\). A call with \code{order = c(x=1, y=0)} is
equivalent to \code{order = c(x=1)}.

To compute numerical derivatives or to evaluate symbolic derivatives at
a point, the function accepts a named vector for the argument
\code{var}; e.g.~\code{var = c(x=1, y=2)} evaluates the derivatives in
\(x=1\) and \(y=2\). For \code{functions} where the first argument is
used as a parameter vector, \code{var} should be a \code{numeric} vector
indicating the point at which the derivatives are to be calculated.

\hypertarget{symbolic-derivatives}{%
\subsection{Symbolic derivatives}\label{symbolic-derivatives}}

Symbolic derivatives are computed via the \code{D} function available in
base \proglang{R}. The function is iterated multiple times for second
and higher order derivatives.

\hypertarget{numerical-derivatives}{%
\subsection{Numerical derivatives}\label{numerical-derivatives}}

Numerical derivatives are computed via the scheme described in
\citet{eberly2008derivative} for central finite differences. In
particular, the derivative of a function \(f\) with respect to one or
more variables is approximated up to the degree \(O(h_1^p\dots h_m^p)\)
by:

\[
\begin{split}
\partial_{n_1,\dots,n_m} f &= 
\partial_{x_1}^{(n_1)}\dots\partial_{x_m}^{(n_m)} f(x_1,\dots,x_m) = \\ 
&=\frac{n_1!\dots n_m!}{h_1^{n_1}\dots h_m^{n_m}} \sum_{j_1=-i^{(n_1)}}^{i^{(n_1)}}\cdots\sum_{j_m=-i^{(n_m)}}^{i^{(n_m)}}C_{j_1}^{(n_1)}\cdots C_{j_m}^{(n_m)}f(x_1+j_1h_1,\dots,x_m+j_mh_m)
\end{split}
\]

where \(n_k\) is the order of differentiation with respect to the
\(k\)-th variable, \(h\) are the step sizes, \(i\) are equal to
\(i^{(n)} = \lfloor(n+p-1)/2\rfloor\), and the coefficients
\(C_j^{(n)}\) are computed by solving the following linear system for
each \(n\):

\[
\begin{bmatrix}
C_{-i}\\
C_{-i+1}\\
C_{-i+2}\\
\vdots\\
C_{-i+n+1}\\
\vdots\\
C_{i}\\
\end{bmatrix} =  
\begin{bmatrix}
(-i)^0 & \dots & (-1)^0 & 0 & 1^0 & \dots & i^0\\
(-i)^1 & \dots & (-1)^1 & 0 & 1^1 & \dots & i^1\\
(-i)^2 & \dots & (-1)^2 & 0 & 1^2 & \dots & i^2\\
\vdots &  & \vdots & \vdots & \vdots &  & \vdots\\
(-i)^{n+1} & \dots & (-1)^{n+1} & 0 & 1^{n+1} & \dots & i^{n+1}\\
\vdots &  & \vdots & \vdots & \vdots &  & \vdots\\
(-i)^{2i} & \dots & (-1)^{2i} & 0 & 1^{2i} & \dots & i^{2i}\\
\end{bmatrix}^{-1}
\begin{bmatrix}
0\\
0\\
0\\
\vdots\\
1\\
\vdots\\
0\\
\end{bmatrix}
\]

The summation is computed via Einstein notation by setting: \[ 
C_{j_1}^{(n_1)}\cdots C_{j_m}^{(n_m)}F_{j_1,\dots,j_m} \equiv \sum_{j_1=-i^{(n_1)}}^{i^{(n_1)}}\cdots\sum_{j_m=-i^{(n_m)}}^{i^{(n_m)}}C_{j_1}^{(n_1)}\cdots C_{j_m}^{(n_m)}f(x_1+j_1h_1,\dots,x_m+j_mh_m)
\]

\hypertarget{examples}{%
\subsection{Examples}\label{examples}}

Symbolic derivatives of univariate functions: \(\partial_x sin(x)\).

\begin{CodeChunk}

\begin{CodeInput}
R> derivative(f = "sin(x)", var = "x")
\end{CodeInput}

\begin{CodeOutput}
[1] "cos(x)"
\end{CodeOutput}
\end{CodeChunk}

Evaluation of symbolic and numerical derivatives:
\(\partial_x sin(x)|_{x=0}\).

\begin{CodeChunk}

\begin{CodeInput}
R> sym <- derivative(f = "sin(x)", var = c(x = 0))
R> num <- derivative(f = function(x) sin(x), var = c(x = 0))
\end{CodeInput}
\end{CodeChunk}

\begin{CodeChunk}

\begin{CodeOutput}
Symbolic  Numeric 
       1        1 
\end{CodeOutput}
\end{CodeChunk}

High order symbolic and numerical derivatives:
\(\partial^{(4)}_x sin(x)|_{x=0}\).

\begin{CodeChunk}

\begin{CodeInput}
R> sym <- derivative(f = "sin(x)", var = c(x = 0), order = 4)
R> num <- derivative(f = function(x) sin(x), var = c(x = 0), order = 4)
\end{CodeInput}
\end{CodeChunk}

\begin{CodeChunk}

\begin{CodeOutput}
  Symbolic    Numeric 
 0.000e+00 -2.922e-11 
\end{CodeOutput}
\end{CodeChunk}

Symbolic derivatives of multivariate functions:
\(\partial_x^{(1)}\partial_y^{(2)} y^2sin(x)\).

\begin{CodeChunk}

\begin{CodeInput}
R> derivative(f = "y^2*sin(x)", var = c("x", "y"), order = c(1, 2))
\end{CodeInput}

\begin{CodeOutput}
[1] "2 * cos(x)"
\end{CodeOutput}
\end{CodeChunk}

Numerical derivatives of multivariate functions:
\(\partial_x^{(1)}\partial_y^{(2)} y^2sin(x)|_{x=0,y=0}\) with degree of
accuracy \(O(h^6)\).

\begin{CodeChunk}

\begin{CodeInput}
R> f <- function(x, y) y^2*sin(x)
R> derivative(f, var = c(x=0, y=0), order = c(1, 2), accuracy = 6)
\end{CodeInput}

\begin{CodeOutput}
[1] 2
\end{CodeOutput}
\end{CodeChunk}

Symbolic gradient of multivariate functions: \(\partial_{x,y}x^2y^2\).

\begin{CodeChunk}

\begin{CodeInput}
R> derivative("x^2*y^2", var = c("x", "y"))
\end{CodeInput}

\begin{CodeOutput}
     [,1]          [,2]           
[1,] "2 * x * y^2" "x^2 * (2 * y)"
\end{CodeOutput}
\end{CodeChunk}

High order derivatives of multivariate functions:
\(\partial^{(6)}_{x,y}x^6y^6\).

\begin{CodeChunk}

\begin{CodeInput}
R> derivative("x^6*y^6", var = c("x", "y"), order = 6)
\end{CodeInput}

\begin{CodeOutput}
     [,1]                            [,2]                             
[1,] "6 * (5 * (4 * (3 * 2))) * y^6" "x^6 * (6 * (5 * (4 * (3 * 2))))"
\end{CodeOutput}
\end{CodeChunk}

Numerical gradient of multivariate functions:
\(\partial_{x,y}x^2y^2|_{x = 1, y = 2}\).

\begin{CodeChunk}

\begin{CodeInput}
R> f <- function(x, y) x^2*y^2
R> derivative(f, var = c(x=1, y=2))
\end{CodeInput}

\begin{CodeOutput}
     [,1] [,2]
[1,]    8    4
\end{CodeOutput}
\end{CodeChunk}

Numerical Jacobian of vector valued functions:
\(\partial_{x,y}[xy,x^2y^2]|_{x = 1, y = 2}\).

\begin{CodeChunk}

\begin{CodeInput}
R> f <- function(x, y) c(x*y, x^2*y^2)
R> derivative(f, var = c(x=1, y=2))
\end{CodeInput}

\begin{CodeOutput}
     [,1] [,2]
[1,]    2    1
[2,]    8    4
\end{CodeOutput}
\end{CodeChunk}

Numerical Jacobian of vector valued \code{functions} where the first
argument is used as a parameter vector:
\(\partial_{X}[\sum_ix_i, \prod_ix_i]|_{X = 0}\).

\begin{CodeChunk}

\begin{CodeInput}
R> f <- function(x) c(sum(x), prod(x))
R> derivative(f, var = c(0, 0, 0))
\end{CodeInput}

\begin{CodeOutput}
     [,1] [,2] [,3]
[1,]    1    1    1
[2,]    0    0    0
\end{CodeOutput}
\end{CodeChunk}

\hypertarget{performance}{%
\subsection{Performance}\label{performance}}

The following table compares the accuracy of Richardson
extrapolation\footnote{\url{https://en.wikipedia.org/wiki/Richardson_extrapolation}}
implemented in the package \pkg{numDeriv} \citep{numDeriv} with central
finite differences implemented in \pkg{calculus} using
\code{accuracy = 4} by default. \(10^4\) derivatives have been computed
for the four functions: \(x^2e^x\), \(xsin(x^2)\), \(xlog(x^2)\),
\(e^{sin(x)}\). The table shows the mean relative error and the
corresponding standard deviation.

\begin{center}
\begin{tabular}{lrrrrrr}
  \toprule
   & pkg &   & N &   & Mean & SD \\ 
    \cmidrule{4-4}  \cmidrule{6-7}
 x$^{2}$ exp(x) & calculus &  & $1.0 \times 10^{4}$ &  & $9.2 \times 10^{-14}$ & $1.0 \times 10^{-11}$ \\ 
   & numDeriv &  & $1.0 \times 10^{4}$ &  & $-3.4 \times 10^{-15}$ & $5.3 \times 10^{-12}$ \\ 
     \cmidrule{4-4}  \cmidrule{6-7}
x sin(x$^{2}$) & calculus &  & $1.0 \times 10^{4}$ &  & $3.8 \times 10^{-13}$ & $1.8 \times 10^{-11}$ \\ 
   & numDeriv &  & $1.0 \times 10^{4}$ &  & $-1.5 \times 10^{-13}$ & $1.2 \times 10^{-11}$ \\ 
     \cmidrule{4-4}  \cmidrule{6-7}
x log(x$^{2}$) & calculus &  & $1.0 \times 10^{4}$ &  & $4.8 \times 10^{-14}$ & $1.3 \times 10^{-11}$ \\ 
   & numDeriv &  & $1.0 \times 10^{4}$ &  & $-4.9 \times 10^{-15}$ & $6.1 \times 10^{-12}$ \\ 
     \cmidrule{4-4}  \cmidrule{6-7}
exp(sin(x)) & calculus &  & $1.0 \times 10^{4}$ &  & $-7.3 \times 10^{-13}$ & $6.0 \times 10^{-11}$ \\ 
   & numDeriv &  & $1.0 \times 10^{4}$ &  & $-1.7 \times 10^{-13}$ & $3.1 \times 10^{-11}$ \\ 
   \bottomrule
\end{tabular}
\end{center}

Although it is known that Richardson extrapolation is usually more
accurate than finite differences, the results of the two packages are
very similar. Both packages produce accurate derivatives with relative
errors close to the precision of the machine (\(10^{-16}\)). On the
other hand, \pkg{calculus} proves to be significantly faster than
\pkg{numDeriv} for multivariate functions as shown in the following
benchmarking.

\begin{CodeChunk}

\begin{CodeInput}
R> x <- rep(0, 1000) 
R> f <- function(x) sum(x)
R> microbenchmark(
R+   "calculus O(2)" = calculus::derivative(f, x, accuracy = 2),
R+   "calculus O(4)" = calculus::derivative(f, x, accuracy = 4),
R+   "calculus O(6)" = calculus::derivative(f, x, accuracy = 6),
R+   "calculus O(8)" = calculus::derivative(f, x, accuracy = 8),
R+   "numDeriv" = numDeriv::grad(f, x)
R+ )
\end{CodeInput}

\begin{CodeOutput}
Unit: milliseconds
          expr     min      lq    mean  median      uq     max neval   cld
 calculus O(2)  17.653  18.546  20.531  20.604  22.260  25.735   100 a    
 calculus O(4)  28.807  30.599  33.137  33.416  34.406  50.392   100  b   
 calculus O(6)  39.758  43.676  49.144  45.171  50.534 156.399   100   c  
 calculus O(8)  51.556  55.662  59.491  56.580  59.303 173.515   100    d 
      numDeriv 102.813 105.561 115.707 108.687 110.493 227.502   100     e
\end{CodeOutput}
\end{CodeChunk}

\hypertarget{taylor-series}{%
\section{Taylor Series}\label{taylor-series}}

\label{sec:taylor}

Based on the derivatives in the previous section, the function
\code{taylor} provides a convenient way to compute the Taylor series of
arbitrary unidimensional or multidimensional functions. The mathematical
function can be specified both as a \code{character} string or as a
\code{function}. Symbolic or numerical methods are applied accordingly.
For univariate functions, the \(n\)-th order Taylor approximation
centered in \(x_0\) is given by:

\[
f(x) \simeq \sum_{k=0}^n\frac{f^{(k)}(x_0)}{k!}(x-x_0)^k
\]

where \(f^{(k)}(x_0)\) denotes the \(k\)-th order derivative evaluated
in \(x_0\). By using multi-index notation, the Taylor series is
generalized to multidimensional functions with an arbitrary number of
variables: \[
f(x) \simeq \sum_{|k|=0}^n\frac{f^{(k)}(x_0)}{k!}(x-x_0)^k
\]

where now \(x=(x_1,\dots,x_d)\) is the vector of variables,
\(k=(k_1,\dots,k_d)\) gives the order of differentiation with respect to
each variable
\(f^{(k)}=\frac{\partial^{(|k|)}f}{\partial^{(k_1)}_{x_1}\cdots \partial^{(k_d)}_{x_d}}\),
and:

\[|k| = k_1+\cdots+k_d \quad\quad k!=k_1!\cdots k_d! \quad\quad x^k=x_1^{k_1}\cdots x_d^{k_d}\]

The summation runs for \(0\leq |k|\leq n\) and identifies the set

\[\{(k_1,\cdots,k_d):k_1+\cdots k_d \leq n\}\]

that corresponds to the partitions of the integer \(n\). These
partitions can be computed with the function \code{partitions} that is
included in the package and optimized in \proglang{C++} for speed and
flexibility. For example, the following call generates the partitions
needed for the \(2\)-nd order Taylor expansion for a function of \(3\)
variables:

\begin{CodeChunk}

\begin{CodeInput}
R> partitions(n = 2, length = 3, fill = TRUE, perm = TRUE, equal = FALSE)
\end{CodeInput}

\begin{CodeOutput}
     [,1] [,2] [,3] [,4] [,5] [,6] [,7] [,8] [,9] [,10]
[1,]    0    0    0    1    0    0    2    0    1     1
[2,]    0    0    1    0    0    2    0    1    0     1
[3,]    0    1    0    0    2    0    0    1    1     0
\end{CodeOutput}
\end{CodeChunk}

Based on these partitions, the function \code{taylor} computes the
corresponding derivatives and builds the Taylor series. The output is a
\code{list} containing the Taylor series, the order of the expansion,
and a \code{data.frame} containing the variables, coefficients and
degrees of each term in the Taylor series.

\begin{CodeChunk}

\begin{CodeInput}
R> taylor("exp(x)", var = "x", order = 2)
\end{CodeInput}

\begin{CodeOutput}
$f
[1] "(1) * 1 + (1) * x^1 + (0.5) * x^2"

$order
[1] 2

$terms
  var coef degree
0   1  1.0      0
1 x^1  1.0      1
2 x^2  0.5      2
\end{CodeOutput}
\end{CodeChunk}

By default, the series is centered in \(x_0=0\) but the function also
supports \(x_0\neq 0\), the multivariable case, and the approximation of
user defined \proglang{R} \code{functions}.

\begin{CodeChunk}

\begin{CodeInput}
R> f <- function(x, y) log(y)*sin(x)
R> taylor(f, var = c(x = 0, y = 1), order = 2)
\end{CodeInput}

\begin{CodeOutput}
$f
[1] "(0.999999999969436) * x^1*(y-1)^1"

$order
[1] 2

$terms
            var coef degree
0,0           1    0      0
0,1     (y-1)^1    0      1
1,0         x^1    0      1
0,2     (y-1)^2    0      2
2,0         x^2    0      2
1,1 x^1*(y-1)^1    1      2
\end{CodeOutput}
\end{CodeChunk}

\hypertarget{hermite-polynomials}{%
\section{Hermite Polynomials}\label{hermite-polynomials}}

\label{sec:hermite}

Hermite polynomials are obtained by differentiation of the Gaussian
kernel:

\[H_{\nu}(x,\Sigma) = exp \Bigl( \frac{1}{2} x_i \Sigma_{ij} x_j \Bigl) (- \partial_x )^\nu exp \Bigl( -\frac{1}{2} x_i \Sigma_{ij} x_j \Bigl)\]

where \(\Sigma\) is a \(d\)-dimensional square matrix and
\(\nu=(\nu_1 \dots \nu_d)\) is the vector representing the order of
differentiation for each variable \(x = (x_1\dots x_d)\). In the case
where \(\Sigma=1\) and \(x=x_1\) the formula reduces to the standard
univariate Hermite polynomials:

\[
H_{\nu}(x) = e^{\frac{x^2}{2}}(-1)^\nu \frac{d^\nu}{dx^\nu}e^{-\frac{x^2}{2}}
\]

High order derivatives of the kernel \(e^{-\frac{x^2}{2}}\) cannot
performed efficiently in base \proglang{R}. The following example shows
the naive calculation of \(\frac{d^2}{dx^2}e^{-\frac{x^2}{2}}\) via the
function \code{D}:

\begin{CodeChunk}

\begin{CodeInput}
R> D(D(expression(exp(-x^2/2)), "x"), "x")
\end{CodeInput}

\begin{CodeOutput}
-(exp(-x^2/2) * (2/2) - exp(-x^2/2) * (2 * x/2) * (2 * x/2))
\end{CodeOutput}
\end{CodeChunk}

The resulting expression is not simplified and this leads to more and
more iterations of the chain rule to compute higher order derivatives.
The expression grows fast and soon requires long computational times and
gigabytes of storage.

\begin{CodeChunk}

\begin{CodeInput}
R> f <- expression(exp(-x^2/2))
R> for(i in 1:14) f <- D(f, "x")
R> object.size(f)
\end{CodeInput}

\begin{CodeOutput}
7925384376 bytes
\end{CodeOutput}
\end{CodeChunk}

To overcome this difficulty, the function \code{hermite} implements the
following scheme. First, it differentiates the gaussian kernel. Then,
the kernel is dropped from the resulting expression. In this way, the
expression becomes a polynomial of degree \(1\). The \code{taylor}
series of order \(1\) is computed in order to extract the coefficients
of the polynomial and rewrite it compact form. The polynomial is now
multiplied by the gaussian kernel and differentiated again. The kernel
is dropped so that the expression becomes a polynomial of degree \(2\).
The \code{taylor} series of order \(2\) is computed and the scheme is
iterated until reaching the desired degree \(\nu\). The same applies
when \(\nu=(\nu_1\dots \nu_d)\) represents the multi index of
multivariate Hermite polynomials. The scheme allows to reduce the
computational time and storage, return a well formatted output, and
generate recursively all the Hermite polynomials of degree \(\nu'\)
where \(|\nu'| \leq|\nu|\). The output is a \code{list} of Hermite
polynomials of degree \(\nu'\), where each polynomial is represented by
the corresponding \code{taylor} series. In the univariate case, for
\(\nu=2\) the Hermite polynomials \(H_0\), \(H_1\), and \(H_2\) are
generated:

\begin{CodeChunk}

\begin{CodeInput}
R> hermite(order = 2)
\end{CodeInput}

\begin{CodeOutput}
$`0`
$`0`$f
[1] "(1) * 1"

$`0`$order
[1] 0

$`0`$terms
  var coef degree
0   1    1      0


$`1`
$`1`$f
[1] "(1) * x^1"

$`1`$order
[1] 1

$`1`$terms
  var coef degree
0   1    0      0
1 x^1    1      1


$`2`
$`2`$f
[1] "(-1) * 1 + (1) * x^2"

$`2`$order
[1] 2

$`2`$terms
  var coef degree
0   1   -1      0
1 x^1    0      1
2 x^2    1      2
\end{CodeOutput}
\end{CodeChunk}

In the multivariate case, where for simplicity
\(\Sigma_{ij}=\delta_{ij}\), \(x=(x_1,x_2)\), and \(|\nu|=2\):

\begin{CodeChunk}

\begin{CodeInput}
R> hermite(order = 2, sigma = diag(2), var = c("x1", "x2"))
\end{CodeInput}

\begin{CodeOutput}
$`0,0`
$`0,0`$f
[1] "(1) * 1"

$`0,0`$order
[1] 0

$`0,0`$terms
    var coef degree
0,0   1    1      0


$`0,1`
$`0,1`$f
[1] "(1) * x2^1"

$`0,1`$order
[1] 1

$`0,1`$terms
     var coef degree
0,0    1    0      0
0,1 x2^1    1      1
1,0 x1^1    0      1


$`1,0`
$`1,0`$f
[1] "(1) * x1^1"

$`1,0`$order
[1] 1

$`1,0`$terms
     var coef degree
0,0    1    0      0
0,1 x2^1    0      1
1,0 x1^1    1      1


$`0,2`
$`0,2`$f
[1] "(-1) * 1 + (1) * x2^2"

$`0,2`$order
[1] 2

$`0,2`$terms
          var coef degree
0,0         1   -1      0
0,1      x2^1    0      1
1,0      x1^1    0      1
0,2      x2^2    1      2
2,0      x1^2    0      2
1,1 x1^1*x2^1    0      2


$`2,0`
$`2,0`$f
[1] "(-1) * 1 + (1) * x1^2"

$`2,0`$order
[1] 2

$`2,0`$terms
          var coef degree
0,0         1   -1      0
0,1      x2^1    0      1
1,0      x1^1    0      1
0,2      x2^2    0      2
2,0      x1^2    1      2
1,1 x1^1*x2^1    0      2


$`1,1`
$`1,1`$f
[1] "(1) * x1^1*x2^1"

$`1,1`$order
[1] 2

$`1,1`$terms
          var coef degree
0,0         1    0      0
0,1      x2^1    0      1
1,0      x1^1    0      1
0,2      x2^2    0      2
2,0      x1^2    0      2
1,1 x1^1*x2^1    1      2
\end{CodeOutput}
\end{CodeChunk}

\hypertarget{ordinary-differential-equations}{%
\section{Ordinary differential
equations}\label{ordinary-differential-equations}}

\label{sec:ode}

The function \code{ode} provides solvers for systems of ordinary
differential equations of the type:

\[
\frac{dy}{dt} = f(t,y), \quad y(t_0)=y_0
\]

where \(y\) is the vector of state variables. Two solvers are available:
the simpler and faster Euler scheme\footnote{\url{https://en.wikipedia.org/wiki/Euler_method}}
or the more accurate 4-th order Runge-Kutta method\footnote{\url{https://en.wikipedia.org/wiki/Runge-Kutta_methods}}.
Although many packages already exist to solve ordinary differential
equations in \proglang{R}\footnote{\url{https://cran.r-project.org/web/views/DifferentialEquations.html}},
they usually represent the function \(f\) either with an \proglang{R}
\code{function} - see e.g.~\pkg{deSolve} \citep{deSolve}, \pkg{odeintr}
\citep{odeintr}, and \pkg{pracma} \citep{pracma} - or with
\code{characters} - see e.g.~\pkg{yuima} \citep{yuima}. While the
representation via \proglang{R} \code{functions} is usually more
efficient, the symbolic representation is easier to adopt for beginners
and more flexible for advanced users to handle systems that might have
been generated via symbolic programming. The package \pkg{calculus}
supports both the representations and uses hashed \code{environments} to
improve symbolic evaluations. Consider the following system:

\[
\frac{d}{dt}
\begin{bmatrix}
x\\
y
\end{bmatrix}=
\begin{bmatrix}
x\\
x(1+\cos(10t))
\end{bmatrix}, \quad
\begin{bmatrix}
x_0\\y_0
\end{bmatrix}=
\begin{bmatrix}
1\\1
\end{bmatrix}
\]

The vector-valued function \(f\) representing the system can be
specified as a vector of \code{characters}, or a \code{function}
returning a numeric vector, giving the values of the derivatives at time
\(t\). The initial conditions are set with the argument \code{var} and
the time variable can be specified with \code{timevar}.

\begin{CodeChunk}

\begin{CodeInput}
R> sim <- ode(f = c("x", "x*(1+cos(10*t))"), 
R+            var = c(x = 1, y = 1), 
R+            times = seq(0, 2*pi, by = 0.001), 
R+            timevar = "t")
\end{CodeInput}
\end{CodeChunk}

\begin{CodeChunk}
\begin{figure}[h]

{\centering \includegraphics{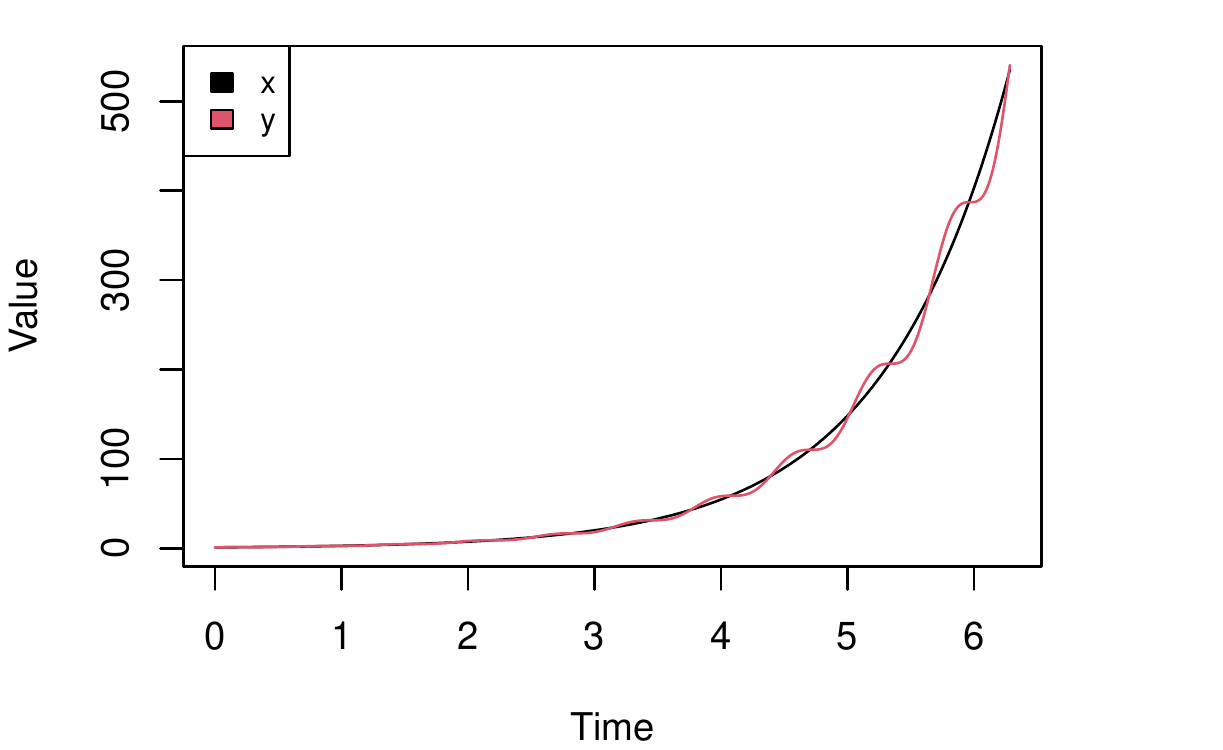} 

}

\caption{Solution to the system of ordinary differential equations in Section \ref{sec:ode} obtained with the function \code{ode} using the Runge-Kutta method.}\label{fig:unnamed-chunk-57}
\end{figure}
\end{CodeChunk}

\hypertarget{differential-operators}{%
\section{Differential operators}\label{differential-operators}}

\label{sec:differentials}

Orthogonal coordinates are a special but extremely common case of
curvilinear coordinates where the coordinate surfaces all meet at right
angles. The chief advantage of non-Cartesian coordinates is that they
can be chosen to match the symmetry of the problem. For example,
spherical coordinates are the most common curvilinear coordinate systems
and are used in Earth sciences, cartography, quantum mechanics,
relativity, and engineering.\footnote{\url{https://en.wikipedia.org/wiki/Curvilinear_coordinates}}
These coordinates may be derived from a set of Cartesian coordinates by
using a transformation that is locally invertible (a one-to-one map) at
each point. This means that one can convert a point given in a Cartesian
coordinate system to its curvilinear coordinates and back. Differential
operators such as the gradient, divergence, curl, and Laplacian can be
transformed from one coordinate system to another via the usage of scale
factors.\footnote{\url{https://en.wikipedia.org/wiki/Orthogonal_coordinates}}
The package implements these operators in Cartesian, polar, spherical,
cylindrical, parabolic coordinates, and supports arbitrary orthogonal
coordinates systems defined by custom scale factors.

\begin{longtable}[]{@{}lll@{}}
\toprule
\begin{minipage}[b]{0.30\columnwidth}\raggedright
Curvilinear coordinates \((q_1, q_2, q_3)\)\strut
\end{minipage} & \begin{minipage}[b]{0.30\columnwidth}\raggedright
Transformation from cartesian \((x, y, z)\)\strut
\end{minipage} & \begin{minipage}[b]{0.30\columnwidth}\raggedright
Scale factors\strut
\end{minipage}\tabularnewline
\midrule
\endhead
\begin{minipage}[t]{0.30\columnwidth}\raggedright
Spherical polar coordinates \((r,\theta ,\phi )\)\strut
\end{minipage} & \begin{minipage}[t]{0.30\columnwidth}\raggedright
\({\begin{aligned}x&=r\sin \theta \cos \phi \\y&=r\sin \theta \sin \phi \\z&=r\cos \theta \end{aligned}}\)\strut
\end{minipage} & \begin{minipage}[t]{0.30\columnwidth}\raggedright
\({\begin{aligned}h_{1}&=1\\h_{2}&=r\\h_{3}&=r\sin \theta \end{aligned}}\)\strut
\end{minipage}\tabularnewline
\begin{minipage}[t]{0.30\columnwidth}\raggedright
---------------\strut
\end{minipage} & \begin{minipage}[t]{0.30\columnwidth}\raggedright
---------------\strut
\end{minipage} & \begin{minipage}[t]{0.30\columnwidth}\raggedright
---------------\strut
\end{minipage}\tabularnewline
\begin{minipage}[t]{0.30\columnwidth}\raggedright
Cylindrical polar coordinates \((r,\phi ,z)\)\strut
\end{minipage} & \begin{minipage}[t]{0.30\columnwidth}\raggedright
\({\begin{aligned}x&=r\cos \phi \\y&=r\sin \phi \\z&=z\end{aligned}}\)\strut
\end{minipage} & \begin{minipage}[t]{0.30\columnwidth}\raggedright
\({\begin{aligned}h_{1}&=h_{3}=1\\h_{2}&=r\end{aligned}}\)\strut
\end{minipage}\tabularnewline
\begin{minipage}[t]{0.30\columnwidth}\raggedright
---------------\strut
\end{minipage} & \begin{minipage}[t]{0.30\columnwidth}\raggedright
---------------\strut
\end{minipage} & \begin{minipage}[t]{0.30\columnwidth}\raggedright
---------------\strut
\end{minipage}\tabularnewline
\begin{minipage}[t]{0.30\columnwidth}\raggedright
Parabolic coordinates \((u,v,\phi )\)\strut
\end{minipage} & \begin{minipage}[t]{0.30\columnwidth}\raggedright
\({\begin{aligned}x&=uv\cos \phi \\y&=uv\sin \phi \\z&={\frac {1}{2}}(u^{2}-v^{2})\end{aligned}}\)\strut
\end{minipage} & \begin{minipage}[t]{0.30\columnwidth}\raggedright
\({\begin{aligned}h_{1}&=h_{2}={\sqrt {u^{2}+v^{2}}}\\h_{3}&=uv\end{aligned}}\)\strut
\end{minipage}\tabularnewline
\begin{minipage}[t]{0.30\columnwidth}\raggedright
---------------\strut
\end{minipage} & \begin{minipage}[t]{0.30\columnwidth}\raggedright
---------------\strut
\end{minipage} & \begin{minipage}[t]{0.30\columnwidth}\raggedright
---------------\strut
\end{minipage}\tabularnewline
\begin{minipage}[t]{0.30\columnwidth}\raggedright
Parabolic cylindrical coordinates \((u,v,z)\)\strut
\end{minipage} & \begin{minipage}[t]{0.30\columnwidth}\raggedright
\({\begin{aligned}x&={\frac {1}{2}}(u^{2}-v^{2})\\y&=uv\\z&=z\end{aligned}}\)\strut
\end{minipage} & \begin{minipage}[t]{0.30\columnwidth}\raggedright
\({\begin{aligned}h_{1}&=h_{2}={\sqrt {u^{2}+v^{2}}}\\h_{3}&=1\end{aligned}}\)\strut
\end{minipage}\tabularnewline
\begin{minipage}[t]{0.30\columnwidth}\raggedright
\strut
\end{minipage} & \begin{minipage}[t]{0.30\columnwidth}\raggedright
\strut
\end{minipage} & \begin{minipage}[t]{0.30\columnwidth}\raggedright
\strut
\end{minipage}\tabularnewline
\bottomrule
\end{longtable}

\hypertarget{gradient}{%
\subsection{Gradient}\label{gradient}}

\label{sec:gradient}

The gradient of a scalar-valued function \(F\) is the vector
\((\nabla F)_i\) whose components are the partial derivatives of \(F\)
with respect to each variable \(i\). In arbitrary orthogonal coordinate
systems, the gradient is expressed in terms of the scale factors \(h_i\)
as follows:

\[(\nabla F)_i = \frac{1}{h_i}\partial_iF\]

The function \code{gradient} implements the symbolic and numeric
gradient of \code{functions}, \code{expressions} and \code{characters}.
In Cartesian coordinates:

\begin{CodeChunk}

\begin{CodeInput}
R> gradient("x*y*z", var = c("x", "y", "z"))
\end{CodeInput}

\begin{CodeOutput}
[1] "y * z" "x * z" "x * y"
\end{CodeOutput}
\end{CodeChunk}

and in spherical coordinates:

\begin{CodeChunk}

\begin{CodeInput}
R> gradient("x*y*z", var = c("x","y","z"), coordinates = "spherical")
\end{CodeInput}

\begin{CodeOutput}
[1] "1/1 * (y * z)"          "1/x * (x * z)"          "1/(x*sin(y)) * (x * y)"
\end{CodeOutput}
\end{CodeChunk}

To support arbitrary orthogonal coordinate systems, it is possible to
pass custom scale factors to the argument \texttt{coordinates}. For
instance, the following call is equivalent to the previous example in
spherical coordinates where the scale factors are now explicitly
specified:

\begin{CodeChunk}

\begin{CodeInput}
R> gradient("x*y*z", var = c("x","y","z"), coordinates = c(1,"x","x*sin(y)"))
\end{CodeInput}

\begin{CodeOutput}
[1] "1/(1) * (y * z)"        "1/(x) * (x * z)"        "1/(x*sin(y)) * (x * y)"
\end{CodeOutput}
\end{CodeChunk}

Numerical methods are applied when working with \code{functions} with
the same sintax introduced for derivatives in section
\ref{sec:derivatives}:

\begin{CodeChunk}

\begin{CodeInput}
R> f <- function(x, y, z) x*y*z
R> gradient(f, var = c(x = 1, y = pi/2, z = 0), coordinates = "spherical")
\end{CodeInput}

\begin{CodeOutput}
[1] 0.0000 0.0000 1.5708
\end{CodeOutput}
\end{CodeChunk}

or in vectorized form:

\begin{CodeChunk}

\begin{CodeInput}
R> f <- function(x) x[1]*x[2]*x[3]
R> gradient(f, var = c(1, pi/2, 0), coordinates = "spherical")
\end{CodeInput}

\begin{CodeOutput}
[1] 0.0000 0.0000 1.5708
\end{CodeOutput}
\end{CodeChunk}

When the function \(F\) is a tensor-valued function
\(F_{d_1,\dots,d_n}\), the gradient is computed for each scalar
component.

\[(\nabla F_{d_1,\dots,d_n})_i = \frac{1}{h_i}\partial_iF_{d_1,\dots,d_n}\]

In particular, this reduces to the Jacobian matrix for vector-valued
functions \(F_{d_1}\):

\begin{CodeChunk}

\begin{CodeInput}
R> f <- function(x) c(prod(x), sum(x))
R> gradient(f, var = c(3, 2, 1))
\end{CodeInput}

\begin{CodeOutput}
     [,1] [,2] [,3]
[1,]    2    3    6
[2,]    1    1    1
\end{CodeOutput}
\end{CodeChunk}

that may be expressed in arbitrary orthogonal coordinate systems.

\begin{CodeChunk}

\begin{CodeInput}
R> f <- function(x) c(prod(x), sum(x))
R> gradient(f, var = c(3, 2, 1), coordinates = "cylindrical")
\end{CodeInput}

\begin{CodeOutput}
     [,1]    [,2] [,3]
[1,]    2 1.00000    6
[2,]    1 0.33333    1
\end{CodeOutput}
\end{CodeChunk}

\hypertarget{jacobian}{%
\subsubsection{Jacobian}\label{jacobian}}

The function \code{jacobian} is a wrapper for \code{gradient} that
always returns the Jacobian as a \code{matrix}, even in the case of
unidimensional scalar-valued functions.

\begin{CodeChunk}

\begin{CodeInput}
R> f <- function(x) x^2
R> jacobian(f, var = c(1))
\end{CodeInput}

\begin{CodeOutput}
     [,1]
[1,]    2
\end{CodeOutput}
\end{CodeChunk}

\hypertarget{hessian}{%
\subsubsection{Hessian}\label{hessian}}

In Cartesian coordinates, the Hessian of a scalar-valued function \(F\)
is the square matrix of second-order partial derivatives:

\[(H(F))_{ij} = \partial_{ij}F\]

It might be tempting to apply the definition of the Hessian as the
Jacobian of the gradient to write it in terms of the scale factors.
However, this results in a Hessian matrix that is not symmetric and
ignores the distinction between vector and covectors in tensor analysis
(see e.g.~\citet{masi2007compressive}). The generalization to arbitrary
coordinate system is out of the scope of this paper and only Cartesian
coordinates are supported:

\begin{CodeChunk}

\begin{CodeInput}
R> f <- function(x, y, z) x*y*z
R> hessian(f, var = c(x = 3, y = 2, z = 1))
\end{CodeInput}

\begin{CodeOutput}
           [,1]       [,2]       [,3]
[1,] 1.2223e-11 1.0000e+00 2.0000e+00
[2,] 1.0000e+00 2.7501e-11 3.0000e+00
[3,] 2.0000e+00 3.0000e+00 1.1001e-10
\end{CodeOutput}
\end{CodeChunk}

When the function \(F\) is a tensor-valued function
\(F_{d_1,\dots,d_n}\), the \code{hessian} is computed for each scalar
component.

\[(H(F_{d_1,\dots,d_n}))_{ij} = \partial_{ij}F_{d_1,\dots,d_n}\]

In this case, the function returns an \code{array} of Hessian matrices:

\begin{CodeChunk}

\begin{CodeInput}
R> f <- function(x, y, z) c(x*y*z, x+y+z)
R> h <- hessian(f, var = c(x = 3, y = 2, z = 1))
\end{CodeInput}
\end{CodeChunk}

that can be extracted with the corresponding indices.

\begin{CodeChunk}

\begin{CodeInput}
R> h[1,,]
\end{CodeInput}

\begin{CodeOutput}
           [,1]       [,2]       [,3]
[1,] 1.2223e-11 1.0000e+00 2.0000e+00
[2,] 1.0000e+00 2.7501e-11 3.0000e+00
[3,] 2.0000e+00 3.0000e+00 1.1001e-10
\end{CodeOutput}

\begin{CodeInput}
R> h[2,,]
\end{CodeInput}

\begin{CodeOutput}
            [,1]        [,2]        [,3]
[1,] -1.8334e-11  7.8835e-12 -3.6273e-12
[2,]  7.8835e-12 -6.4170e-11  1.0907e-11
[3,] -3.6273e-12  1.0907e-11  9.1671e-11
\end{CodeOutput}
\end{CodeChunk}

\hypertarget{divergence}{%
\subsection{Divergence}\label{divergence}}

The divergence of a vector-valued function \(F_i\) produces a scalar
value \(\nabla \cdot F\) representing the volume density of the outward
flux of the vector field from an infinitesimal volume around a given
point.\footnote{\url{https://en.wikipedia.org/wiki/Divergence}} In terms
of scale factors, it is expressed as follows:

\[\nabla \cdot F = \frac{1}{J}\sum_i\partial_i\Biggl(\frac{J}{h_i}F_i\Biggl)\]

where \(J=\prod_ih_i\). When \(F\) is an \code{array} of vector-valued
functions \(F_{d_1,\dots,d_n,i}\), the \code{divergence} is computed for
each vector:

\[(\nabla \cdot F)_{d_1,\dots,d_n} = \frac{1}{J}\sum_i\partial_i\Biggl(\frac{J}{h_i}F_{d_1,\dots,d_n,i}\Biggl)=\frac{1}{J}\sum_i\partial_i(Jh_i^{-1})F_{d_1,\dots,d_n,i}+Jh_i^{-1}\partial_i(F_{d_1,\dots,d_n,i})\]

where the last equality is preferable in practice as the derivatives of
the scale factor can be computed symbolically and the computation of the
derivatives of \(F\) is more efficient than the direct computation of
\(\partial_i\bigl(\frac{J}{h_i}F_{d_1,\dots,d_n,i}\bigl)\) via finite
differences. In Cartesian coordinates:

\begin{CodeChunk}

\begin{CodeInput}
R> f <- c("x^2", "y^2", "z^2")
R> divergence(f, var = c("x","y","z"))
\end{CodeInput}

\begin{CodeOutput}
[1] "2 * x + 2 * y + 2 * z"
\end{CodeOutput}
\end{CodeChunk}

In polar coordinates:

\begin{CodeChunk}

\begin{CodeInput}
R> f <- c("sqrt(r)/10", "sqrt(r)")
R> divergence(f, var = c("r","phi"), coordinates = "polar")
\end{CodeInput}

\begin{CodeOutput}
[1] "(0.5 * r^-0.5/10 * r + (sqrt(r)/10)) / (1*r)"
\end{CodeOutput}
\end{CodeChunk}

And for tensors of vector-valued functions:

\begin{CodeChunk}

\begin{CodeInput}
R> f <- matrix(c("x^2","y^2","z^2","x","y","z"), nrow = 2, byrow = TRUE)
R> divergence(f, var = c("x","y","z"))
\end{CodeInput}

\begin{CodeOutput}
[1] "2 * x + 2 * y + 2 * z" "1 + 1 + 1"            
\end{CodeOutput}
\end{CodeChunk}

The same syntax holds for \code{functions} where numerical methods are
automatically applied:

\begin{CodeChunk}

\begin{CodeInput}
R> f <- function(x,y,z) matrix(c(x^2,y^2,z^2,x,y,z), nrow = 2, byrow = TRUE)
R> divergence(f, var = c(x = 0, y = 0, z = 0))
\end{CodeInput}

\begin{CodeOutput}
[1] 0 3
\end{CodeOutput}
\end{CodeChunk}

\hypertarget{curl}{%
\subsection{Curl}\label{curl}}

The curl of a vector-valued function \(F_i\) at a point is represented
by a vector whose length and direction denote the magnitude and axis of
the maximum circulation.\footnote{\url{https://en.wikipedia.org/wiki/Curl_(mathematics)}}
In 2 dimensions, the curl is written in terms of the scale factors \(h\)
and the Levi-Civita symbol \(\epsilon\) as follows:

\[\nabla \times F = \frac{1}{h_1h_2}\sum_{ij}\epsilon_{ij}\partial_i\Bigl(h_jF_j\Bigl)= \frac{1}{h_1h_2}\Biggl(\partial_1\Bigl(h_2F_2\Bigl)-\partial_2\Bigl(h_1F_1\Bigl)\Biggl)\]

In 3 dimensions:

\[(\nabla \times F)_k = \frac{h_k}{J}\sum_{ij}\epsilon_{ijk}\partial_i\Bigl(h_jF_j\Bigl)\]

where \(J=\prod_i h_i\). This suggests to implement the \code{curl} in
\(m+2\) dimensions in such a way that the formula reduces correctly to
the previous cases:

\[(\nabla \times F)_{k_1\dots k_m} = \frac{h_{k_1}\cdots h_{k_m}}{J}\sum_{ij}\epsilon_{ijk_1\dots k_m}\partial_i\Bigl(h_jF_j\Bigl)\]

And in particular, when \(F\) is an \code{array} of vector-valued
functions \(F_{d_1,\dots,d_n,i}\) the \code{curl} is computed for each
vector:

\[
\begin{split}
(\nabla \times F)_{d_1\dots d_n,k_1\dots k_m} & =\\ 
&=\frac{h_{k_1}\cdots h_{k_m}}{J}\sum_{ij}\epsilon_{ijk_1\dots k_m}\partial_i\Bigl(h_jF_{d_1\dots d_n,j}\Bigl) \\
&=\sum_{ij}\frac{1}{h_ih_j}\epsilon_{ijk_1\dots k_m}\partial_i\Bigl(h_jF_{d_1\dots d_n,j}\Bigl) \\
&=\sum_{ij}\frac{1}{h_ih_j}\epsilon_{ijk_1\dots k_m}\Bigl(\partial_i(h_j)F_{d_1\dots d_n,j}+h_j\partial_i(F_{d_1\dots d_n,j})\Bigl)
\end{split}
\]

where the last equality is preferable in practice as the derivatives of
the scale factor can be computed symbolically and the computation of the
derivatives of \(F\) is more efficient than the direct computation of
\(\partial_i\bigl(h_jF_{d_1\dots d_n,j}\bigl)\) via finite differences.
In 2-dimensional Cartesian coordinates:

\begin{CodeChunk}

\begin{CodeInput}
R> f <- c("x^3*y^2","x")
R> curl(f, var = c("x","y"))
\end{CodeInput}

\begin{CodeOutput}
[1] "(1) * 1 + (x^3 * (2 * y)) * -1"
\end{CodeOutput}
\end{CodeChunk}

In 3 dimensions, for an irrotational vector field:

\begin{CodeChunk}

\begin{CodeInput}
R> f <- c("x","-y","z")
R> curl(f, var = c("x","y","z"))
\end{CodeInput}

\begin{CodeOutput}
[1] "0" "0" "0"
\end{CodeOutput}
\end{CodeChunk}

And for tensors of vector-valued functions:

\begin{CodeChunk}

\begin{CodeInput}
R> f <- matrix(c("x","-y","z","x^3*y^2","x","0"), nrow = 2, byrow = TRUE)
R> curl(f, var = c("x","y","z"))
\end{CodeInput}

\begin{CodeOutput}
     [,1] [,2] [,3]                            
[1,] "0"  "0"  "0"                             
[2,] "0"  "0"  "(1) * 1 + (x^3 * (2 * y)) * -1"
\end{CodeOutput}
\end{CodeChunk}

The same syntax holds for \code{functions} where numerical methods are
automatically applied and for arbitrary orthogonal coordinate systems as
shown in the previous sections.

\hypertarget{laplacian}{%
\subsection{Laplacian}\label{laplacian}}

The Laplacian is a differential operator given by the divergence of the
gradient of a scalar-valued function \(F\), resulting in a scalar value
giving the flux density of the gradient flow of a function. The
Laplacian occurs in differential equations that describe many physical
phenomena, such as electric and gravitational potentials, the diffusion
equation for heat and fluid flow, wave propagation, and quantum
mechanics.\footnote{\url{https://en.wikipedia.org/wiki/Laplace_operator}}
In terms of the scale factor, the operator is written as:

\[\nabla^2F = \frac{1}{J}\sum_i\partial_i\Biggl(\frac{J}{h_i^2}\partial_iF\Biggl)\]

where \(J=\prod_ih_i\). When the function \(F\) is a tensor-valued
function \(F_{d_1,\dots,d_n}\), the \code{laplacian} is computed for
each scalar component:

\[(\nabla^2F)_{d_1\dots d_n} = \frac{1}{J}\sum_i\partial_i\Biggl(\frac{J}{h_i^2}\partial_iF_{d_1\dots d_n}\Biggl)=\frac{1}{J}\sum_i\partial_i\Bigl(Jh_i^{-2}\Bigl)\partial_iF_{d_1\dots d_n}+Jh_i^{-2}\partial^2_iF_{d_1\dots d_n}\]

where the last equality is preferable in practice as the derivatives of
the scale factor can be computed symbolically and the computation of the
derivatives of \(F\) is more efficient than the direct computation of
\(\partial_i\bigl(\frac{J}{h_i^2}\partial_iF\bigl)\) via finite
differences. In Cartesian coordinates:

\begin{CodeChunk}

\begin{CodeInput}
R> f <- "x^3+y^3+z^3"
R> laplacian(f, var = c("x","y","z"))
\end{CodeInput}

\begin{CodeOutput}
[1] "3 * (2 * x) + 3 * (2 * y) + 3 * (2 * z)"
\end{CodeOutput}
\end{CodeChunk}

And for tensors of scalar-valued functions:

\begin{CodeChunk}

\begin{CodeInput}
R> f <- array(c("x^3+y^3+z^3", "x^2+y^2+z^2", "y^2", "z*x^2"), dim = c(2,2))
R> laplacian(f, var = c("x","y","z"))
\end{CodeInput}

\begin{CodeOutput}
     [,1]                                      [,2]   
[1,] "3 * (2 * x) + 3 * (2 * y) + 3 * (2 * z)" "2"    
[2,] "2 + 2 + 2"                               "z * 2"
\end{CodeOutput}
\end{CodeChunk}

The same syntax holds for \code{functions} where numerical methods are
automatically applied and for arbitrary orthogonal coordinate systems as
shown in the previous sections.

\hypertarget{integrals}{%
\section{Integrals}\label{integrals}}

\label{sec:integrals}

The package integrates seamlessly with \pkg{cubature} \citep{cubature}
for efficient numerical integration in \proglang{C}. The function
\code{integral} provides the interface for multidimensional integrals of
\code{functions}, \code{expressions}, and \code{characters} in arbitrary
orthogonal coordinate systems. If the package \pkg{cubature} is not
installed, the package implements a naive Monte Carlo integration by
default. The function returns a \code{list} containing the \code{value}
of the integral as well as other information on the estimation
uncertainty. The integration bounds are specified via the argument
\code{bounds}: a list containing the lower and upper bound for each
variable. If the two bounds coincide, or if a single number is
specified, the corresponding variable is not integrated and its value is
fixed. For arbitrary orthogonal coordinates \(q_1\dots q_n\) the
integral is computed as:

\[
\int J\cdot f(q_1\dots q_n) dq_1\dots dq_n
\]

where \(J=\prod_i h_i\) is the Jacobian determinant of the
transformation and is equal to the product of the scale factors
\(h_1\dots h_n\).

\hypertarget{examples-1}{%
\subsection{Examples}\label{examples-1}}

Univariate integral \(\int_0^1xdx\):

\begin{CodeChunk}

\begin{CodeInput}
R> i <- integral(f = "x", bounds = list(x = c(0,1)))
R> i$value
\end{CodeInput}

\begin{CodeOutput}
[1] 0.5
\end{CodeOutput}
\end{CodeChunk}

that is equivalent to:

\begin{CodeChunk}

\begin{CodeInput}
R> i <- integral(f = function(x) x, bounds = list(x = c(0,1)))
R> i$value
\end{CodeInput}

\begin{CodeOutput}
[1] 0.5
\end{CodeOutput}
\end{CodeChunk}

Univariate integral \(\int_0^1yxdx|_{y=2}\):

\begin{CodeChunk}

\begin{CodeInput}
R> i <- integral(f = "y*x", bounds = list(x = c(0,1), y = 2))
R> i$value
\end{CodeInput}

\begin{CodeOutput}
[1] 1
\end{CodeOutput}
\end{CodeChunk}

Multivariate integral \(\int_0^1\int_o^1yxdxdy\):

\begin{CodeChunk}

\begin{CodeInput}
R> i <- integral(f = "y*x", bounds = list(x = c(0,1), y = c(0,1)))
R> i$value
\end{CodeInput}

\begin{CodeOutput}
[1] 0.25
\end{CodeOutput}
\end{CodeChunk}

Area of a circle \(\int_0^{2\pi}\int_0^1dA(r,\theta)\)

\begin{CodeChunk}

\begin{CodeInput}
R> i <- integral(f = 1, 
R+               bounds = list(r = c(0,1), theta = c(0,2*pi)), 
R+               coordinates = "polar")
R> i$value
\end{CodeInput}

\begin{CodeOutput}
[1] 3.1416
\end{CodeOutput}
\end{CodeChunk}

Volume of a sphere \(\int_0^\pi\int_0^{2\pi}\int_0^1dV(r,\theta,\phi)\)

\begin{CodeChunk}

\begin{CodeInput}
R> i <- integral(f = 1, 
R+               bounds = list(r = c(0,1), theta = c(0,pi), phi = c(0,2*pi)), 
R+               coordinates = "spherical")
R> i$value
\end{CodeInput}

\begin{CodeOutput}
[1] 4.1888
\end{CodeOutput}
\end{CodeChunk}

As a final example consider the electric potential in spherical
coordinates \(V=\frac{1}{4\pi r}\) arising from a unitary point charge:

\begin{CodeChunk}

\begin{CodeInput}
R> V <- "1/(4*pi*r)"
\end{CodeInput}
\end{CodeChunk}

The electric field is determined by the gradient of the
potential\footnote{\url{https://en.wikipedia.org/wiki/Electric_potential}}
\(E = -\nabla V\):

\begin{CodeChunk}

\begin{CodeInput}
R> E <- -1 %prod% gradient(V, c("r","theta","phi"), coordinates = "spherical")
\end{CodeInput}
\end{CodeChunk}

Then, by Gauss's law\footnote{\url{https://en.wikipedia.org/wiki/Gauss\%27s_law}},
the total charge enclosed within a given volume is equal to the surface
integral of the electric field \(q=\int E\cdot dA\) where \(\cdot\)
denotes the scalar product between the two vectors. In spherical
coordinates, this reduces to the surface integral of the radial
component of the electric field \(\int E_rdA\). The following code
computes this surface integral on a sphere with fixed radius \(r=1\):

\begin{CodeChunk}

\begin{CodeInput}
R> i <- integral(E[1], 
R+               bounds = list(r = 1, theta = c(0,pi), phi = c(0,2*pi)), 
R+               coordinates = "spherical")
R> i$value
\end{CodeInput}

\begin{CodeOutput}
[1] 1
\end{CodeOutput}
\end{CodeChunk}

As expected \(q=\int E\cdot dA=\int E_rdA=1\), the unitary charge
generating the electric potential.

\hypertarget{summary}{%
\section{Summary}\label{summary}}

\label{sec:summary}

This work has presented the \pkg{calculus} package for high dimensional
numerical and symbolic calculus in \proglang{R}. The library applies
numerical methods when working with \code{functions} or symbolic
programming when working with \code{characters} or \code{expressions}.
To describe multidimensional objects such as vectors, matrices, and
tensors, the package uses the class \code{array} regardless of the
dimension. This is done to prevent unwanted results due to operations
among different classes such as \code{vector} for unidimensional objects
or \code{matrix} for bidimensional objects.

The package handles multivariate numerical calculus in arbitrary
dimensions and coordinates via \proglang{C++} optimized functions. It
achieves approximately the same accuracy for numerical differentiation
as the \pkg{numDeriv} \citep{numDeriv} package but significantly reduces
the computational time. It supports higher order derivatives and the
differentiation of possibly tensor-valued functions. Differential
operators such as the gradient, divergence, curl, and Laplacian are made
available in arbitrary orthogonal coordinate systems. The Einstein
summing convention supports expressions involving more than two tensors
and tensors with repeated indices. Besides being more flexible, the
summation proves to be faster than the alternative implementation found
in the \pkg{tensorA} \citep{tensorA} package for advanced tensor
arithmetic with named indices. Unlike \pkg{mpoly} \citep{kahle2013mpoly}
and \pkg{pracma} \citep{pracma}, the package supports multidimensional
Hermite polynomials and Taylor series of multivariate functions. The
package integrates seamlessly with \pkg{cubature} \citep{cubature} for
efficient numerical integration in \proglang{C} and extends the
numerical integration to arbitrary orthogonal coordinate systems.

The symbolic counterpart of the numerical methods are implemented
whenever possible to meet the growing needs for \proglang{R} to handle
basic symbolic operations. The package provides, among others, symbolic
high order derivatives of possibly tensor-valued functions, symbolic
differential operators such as the gradient, divergence, curl, and
Laplacian in arbitrary orthogonal coordinate systems, symbolic Einstein
summing convention and Taylor series expansion of multivariate
functions. This is done entirely in \proglang{R}, without depending on
external computer algebra systems.

Except for \pkg{Rcpp} \citep{eddelbuettel2011rcpp}, the \pkg{calculus}
package has no strict dependencies in order to provide a stable
self-contained toolbox that invites re-use.

\hypertarget{computational-details}{%
\section{Computational details}\label{computational-details}}

The results in this paper were obtained using R 4.0.0 \citep{r} with the
packages \pkg{numDeriv} 2016.8-1.1 \citep{numDeriv}, \pkg{tensorA}
0.36.2 \citep{tensorA}, \pkg{cubature} 2.0.4 \citep{cubature},
\pkg{microbenchmark} 1.4-7 \citep{microbenchmark}, \pkg{calculus} 0.3.0.
\proglang{R} itself and all packages used are available from the
Comprehensive R Archive Network (CRAN) at
\url{https://CRAN.R-project.org/}.

\renewcommand\refname{References}
\bibliography{bibliography.bib}

\end{document}